\documentclass[10pt,a4paper, twocolumn]{article}
\usepackage[a4paper, left=1in, right=1in, top=1in, bottom=1in, columnsep=20pt]{geometry}
\usepackage[english]{babel}
\usepackage{hyphenat}
\usepackage{microtype}
\usepackage{titlesec}

\titleformat{\section}{\normalfont\normalsize\bfseries} 
  {\thesection}{1em}{}
\titleformat{\subsection}
{\normalfont\small\bfseries}
  {\thesubsection}{1em}{} 

\usepackage{bold-extra}
\usepackage{amsmath}
\usepackage{amssymb}
\usepackage{mathtools} 
\usepackage{braket} 
\DeclarePairedDelimiter\ceil{\lceil}{\rceil}
\usepackage{graphicx}
\usepackage{appendix}
\usepackage{authblk}
\usepackage{csquotes}
\usepackage{xcolor}
\usepackage{etoolbox}
\usepackage{balance}

\usepackage{comment}

\pdfoutput=1
\usepackage{color}
\usepackage{todonotes}
\usepackage{subfigure}
\usepackage{booktabs}
\usepackage{array}
\usepackage{diagbox}
\usepackage{yfonts}
\usepackage{multirow,rotating}
\usepackage{amsmath,amssymb,bm,graphicx,wasysym,mathrsfs,dsfont}
\usepackage{cite}
\usepackage{slashed,relsize}
\usepackage[colorlinks=true,linkcolor=black,anchorcolor=black,citecolor=blue,filecolor=cyan,menucolor=red,runcolor=filecolor,urlcolor=blue,bookmarks=true,bookmarksnumbered=true]{hyperref}

\usepackage[font=small,labelfont=bf]{caption}

\def\beq{\begin{equation}\displaystyle\displaystyle}
	\def\eeq{\end{equation}}
\def\bea{\begin{eqnarray}\displaystyle} 
	\def\eea{\end{eqnarray}}

\def\({\left(}
\def\){\right)}
\def\bry{\begin{array}}
	\def\ery{\end{array}}

\makeatletter
\DeclareFontFamily{OMX}{MnSymbolE}{}
\DeclareSymbolFont{MnLargeSymbols}{OMX}{MnSymbolE}{m}{n}
\SetSymbolFont{MnLargeSymbols}{bold}{OMX}{MnSymbolE}{b}{n}
\DeclareFontShape{OMX}{MnSymbolE}{m}{n}{
    <-6>  MnSymbolE5
   <6-7>  MnSymbolE6
   <7-8>  MnSymbolE7
   <8-9>  MnSymbolE8
   <9-10> MnSymbolE9
  <10-12> MnSymbolE10
  <12->   MnSymbolE12
}{}
\DeclareFontShape{OMX}{MnSymbolE}{b}{n}{
    <-6>  MnSymbolE-Bold5
   <6-7>  MnSymbolE-Bold6
   <7-8>  MnSymbolE-Bold7
   <8-9>  MnSymbolE-Bold8
   <9-10> MnSymbolE-Bold9
  <10-12> MnSymbolE-Bold10
  <12->   MnSymbolE-Bold12
}{}

\let\llangle\@undefined
\let\rrangle\@undefined
\DeclareMathDelimiter{\llangle}{\mathopen}%
                     {MnLargeSymbols}{'164}{MnLargeSymbols}{'164}
\DeclareMathDelimiter{\rrangle}{\mathclose}%
                     {MnLargeSymbols}{'171}{MnLargeSymbols}{'171}
\makeatletter

\patchcmd{\AB@authlist}{,}{\unskip~}{}{}
\makeatother

\usepackage{geometry}
\title{Harnessing Quantum Extreme Learning Machines\\ for image classification}
\author[1, 2, *]{A. De Lorenzis}
\author[1, 3]{\protect\mbox{M. P. Casado}}
\author[2]{\protect\mbox{M. P. Estarellas}}
\author[4, 5]{N. Lo Gullo}
\author[1]{T. Lux}
\author[4, 5]{\protect\mbox{F. Plastina}}
\author[3]{\protect\mbox{A. Riera}}
\author[4, 5, 6]{J. Settino}
\affil[1]{Institut de Física d’Altes Energies (IFAE) - The Barcelona Institute of Science and Technology (BIST), Campus UAB, 08193 Bellaterra (Barcelona), Spain}
\affil[2]{Qilimanjaro Quantum Tech S.L., Carrer de Veneçuela, 74, 08019 Barcelona, Spain }
\affil[3]{Departament de Física, Universitat Autònoma de Barcelona }
\affil[4]{Dipartimento di Fisica, Università della Calabria, 87036 Arcavacata di Rende (CS), Italy}
\affil[5]{INFN, gruppo collegato di Cosenza, 87036 Arcavacata di Rende (CS), Italy}
\affil[6]{ICAR-CNR, 87036 Rende (CS), Italy}
\date{}
\begin{document}
\twocolumn[
    \begin{@twocolumnfalse}
    \maketitle
    \centering
    \vspace{-1cm}
    {\scriptsize{* E-mail: \texttt{adelorenzis@ifae.es, \\
    \tiny All the authors, except for the first, are listed in alphabetic order.}}}
    \noindent\line(2,0){420}\\
    \begin{abstract}
        Interest in quantum machine learning is increasingly growing due to its potential to offer more efficient solutions for problems that are difficult to tackle with classical methods. In this context, the research work presented here focuses on the use of quantum machine learning techniques for image classification tasks. We exploit a quantum extreme learning machine by taking advantage of its rich feature map provided by the quantum reservoir substrate.
        We 
        systematically analyze different phases of the quantum extreme learning machine process, from the dataset preparation to the image final classification. In particular, we have tested different encodings, together with Principal Component Analysis, the use of Auto-Encoders, as well as the dynamics of the model through the use of different Hamiltonians for the quantum reservoir.
        Our results show that the introduction of a quantum reservoir systematically improves the accuracy of the classifier. Additionally, our findings indicate that variations in encoding methods can significantly influence performance and that Hamiltonians with distinct structures exhibit the same discrimination rate, depending on how their eigenstates are related to the encoding and measurement basis.
    \end{abstract}
    \noindent\line(2,0){420}\\
    \vspace{1cm}
    \end{@twocolumnfalse}
]
\thispagestyle{empty}
\setcounter{equation}{0}
\setcounter{footnote}{0}
\setcounter{page}{1}
\section{Introduction}\label{sec:intro}
Since the early 2000s, quantum computing has steadily gained prominence, emerging as a groundbreaking technology with the potential to revolutionize various sectors, including chemistry, cryptography, finance, and artificial intelligence \cite{feynman1982simulating, Shordoi:10.1137/S0097539795293172, Montanaro_2016, Nielsen_Chuang_2010}. In recent years, significant technological advancements \cite{Arute48651, Preskill:2018jim, Huang:2021pei, Zhong:2020iql} have accelerated interest in this field, bringing quantum computing closer to practical applications.\\
One particularly exciting and rapidly evolving domain within quantum computing is quantum machine learning (QML), which lies at the intersection of quantum computing and machine learning \cite{Biamonte:2016ugo, book, Schuld:2018gao, Congarticle, Liu:2022bpb}. QML aims to leverage quantum algorithms and data structures to tackle complex problems and achieve breakthroughs that classical machine learning methods struggle to address \cite{Havlicek:2018nqz, Lloyd:2013lby,Schuld2014,mastroianni2023assessing,mastroianni2024variational,consiglio2024variational}.\\
In this paper, we explore the Quantum Extreme Learning Machine (QELM) \cite{article, Innocenti2023, Sakurai:2022ala, Xiong:2023qqh, Kornjaca2024, chenPhysRevApplied.14.024065, QELMSuprano, Vetrano:2024vbh}, a model that utilizes quantum dynamics to perform complex data transformations. The QELM builds on the Extreme Learning Machine (ELM) \cite{1380068,10.1007/s10462-013-9405-z,HUANG201532, HUANG2006489, articleHuang2011, wang2022review, markowska-kaczmar_extreme_2021} framework, a classical machine learning model designed to streamline the training process by limiting learning to the output layer. In the QELM, this approach is extended into the quantum realm, where a fixed-structure quantum reservoir processes input data and maps it into a high-dimensional space, yielding rich representations of the data.\\
Unlike classical Extreme Learning Machines, the QELM harnesses the vast state space of quantum systems, enabling it to handle complex datasets with computational power and efficiency that surpasses traditional methods \cite{Schuld2014}. Additionally, while classical ELMs are constrained by classical linear algebra, the QELM exploits the high dimensionality of quantum systems, making it highly effective for pattern recognition and classification tasks \cite{Biamonte:2016ugo}.\\
The input data is first encoded into quantum states using specific encoding schemes, which are then fed into a quantum reservoir. This reservoir (a system with fixed internal dynamics) transforms the input into higher-dimensional representations where a larger distinguishability is possible. Importantly, the quantum reservoir operates without requiring optimization or training of its internal parameters, allowing complex data transformations to occur naturally. Once processed, the data is passed to the output layer, where learning occurs. Only the weights of this layer are optimized, which significantly reduces training complexity and computational cost.
One of the primary advantages of the QELM is its efficiency in training. Unlike many machine learning models that rely on backpropagation and iterative adjustment of all parameters, the QELM confines learning to the output layer. This approach greatly reduces computational overhead, resulting in faster training times and lower resource consumption.\\
Furthermore, the quantum reservoir's ability to map input data into a vast Hilbert space, which grows exponentially with the number of quantum particles, allows the QELM to capture intricate relationships within the data, revealing patterns that classical methods might miss. This ability to handle complex, high-dimensional transformations is particularly valuable in domains where feature extraction and pattern recognition are key challenges.\\
QELMs fall within the broader framework of Quantum Reservoir Computing (QRC) \cite{Fujii2017, chenPhysRevApplied.14.024065, AngelatosPhysRevX.11.041062}, but operate without memory, i.e. the reservoir's dynamics are reduced to a static, high-dimensional input transformation \cite{Mujal_2021}. While classical reservoir computing has been extensively studied in the context of recurrent neural networks (RNNs) \cite{Jaeger2004,Maass2002,Ma2023,Gauthier2021,Xiao2021, lukovsevivcius2009reservoir, jaeger2001echo} and physical systems \cite{appeltant2011information, van2017advances, torrejon2017neuromorphic, tanaka2019recent}, several advances have extended this framework to quantum systems \cite{Fujii2017,Mujal2021,Ghosh2021}. In particular, spin chains \cite{Xia2023,Sannia2024,Martínez2021,Gotting2023,Martínez2023,Mujal2022, nakajima2019boosting, chen2019learning, tran2020higher, martinez2020information}, fermionic and bosonic systems \cite{Ghosh2019,HoanTran2023,Llodra2023, ghosh2020universal, ghosh2019neuromorphic, nokkala2020gaussian, govia2020quantum}, quantum oscillators \cite{Govia2021,Dudas2023}, and Rydberg atoms\cite{Henriet2020,Bravo2022,Kornjaca2024} have been explored as quantum physical implementations. Furthermore, to enhance the memory of previous input injections, hybrid approaches and feedback mechanisms have been proposed \cite{Settino2024,Pfeffer2022,Wudarski2023,Ahmed2024,Ahmed2024a,Kobayashi2024a}.
Since QELMs do not retain information from previous inputs, they are less suitable for tasks requiring temporal dependencies or sequence learning, such as time-series forecasting.
{\color{red} 
}\\
However, this memoryless nature simplifies the model’s structure, enabling it to focus on instantaneous data transformations, which is ideal for tasks where temporal dependencies are not essential.
Given its properties, the QELM is particularly well-suited for classification, regression, and pattern recognition tasks in domains that do not rely on time-dependency. In fields such as image recognition, fraud detection, and static data analysis, the ability to efficiently map input data to high-dimensional spaces using quantum resources can lead to significant performance improvements \cite{Sedykh,Biamonte:2016ugo, Schuld2014, Farhi:2018nhu, vapnik74theory}. Furthermore, the rapid training process of the QELM makes it an attractive option for applications where time or computational resources are limited.\\
Throughout this paper, we study the architecture and performance of QELM for image classification tasks. Due to the limitations arising from the reduced number of qubits available today and the small decoherence times of current noisy quantum devices \cite{Preskill:2018jim, Domingo:2022fre}, typical datasets, characterized by a large number of features, cannot be directly fed to the quantum reservoirs and must necessarily be compressed into a lower-dimensional representation. In the context of QELM, this is usually achieved by linear reduction algorithms, such as Principal Component Analysis (PCA).\\
In this work, we perform a step forward and exploit the more complex representation power of nonlinear neural network architectures such as the Auto-Encoders. In particular, we quantitatively show that the reservoir performances are substantially improved when the Auto-Encoder, rather than PCA, is used to compress the number of features.\\
Furthermore, since it is necessary to convert classical information into quantum states that the quantum system can process, we explore various encoding methods, identifying those that work better.
Finally, we compare the results obtained from different Hamiltonians that generate the natural dynamics of the reservoir, gaining some insights into the learning processes and performances of the various reservoirs.
\\
The paper is structured as follows: Section \ref{sec:The model} provides a description of the model used. Sections \ref{sec:Principal Component Analysis vs Auto-Encoders} and \ref{sec:Different encodings} cover techniques for reducing data complexity and methods for encoding classical input information, respectively. In Section \ref{sec:Different Hamiltonians}, we present different Hamiltonian models. Section \ref{sec:ML classifier} details the measurement process and the classical machine learning classifier. Finally, Section \ref{sec:Results} discusses the model's performance and results, followed by the conclusion.\\
To assess the performance of our algorithms, we consider, as a simple classification task, the discrimination of the ten handwritten numbers in the MNIST dataset \cite{mnist}. The latter consists of 70000 pictures of digits, between 0 and 9, stored into grayscale images of 28 $\times$ 28 pixels.\\
To demonstrate that our findings are not limited to a specific dataset and are generally applicable, we also test our approach on the more complex Fashion-MNIST dataset \cite{fashionmnistnovelimagedataset}, for which the same general considerations hold.\\
Furthermore, the accuracies obtained in our study are comparable to or exceed those of other image classification methods based on quantum machine learning, such as Quantum Capsule Networks \cite{Liu:2022bpb}, Quantum Convolutional Neural Networks \cite{QCNNHur:2021zyz}, the QRC proposed by \cite{Kornjaca2024}, the Quantum Bayes Classifiers proposed in \cite{BayesWang:2024atn} or the approaches proposed in \cite{Slabbert:2024fjv, Sein:2024ael}.\\
\section{Overview of Quantum Extreme Learning Machines}\label{sec:The model}
\begin{figure*}[htbp]
    \centering
\includegraphics[width=0.95\textwidth]{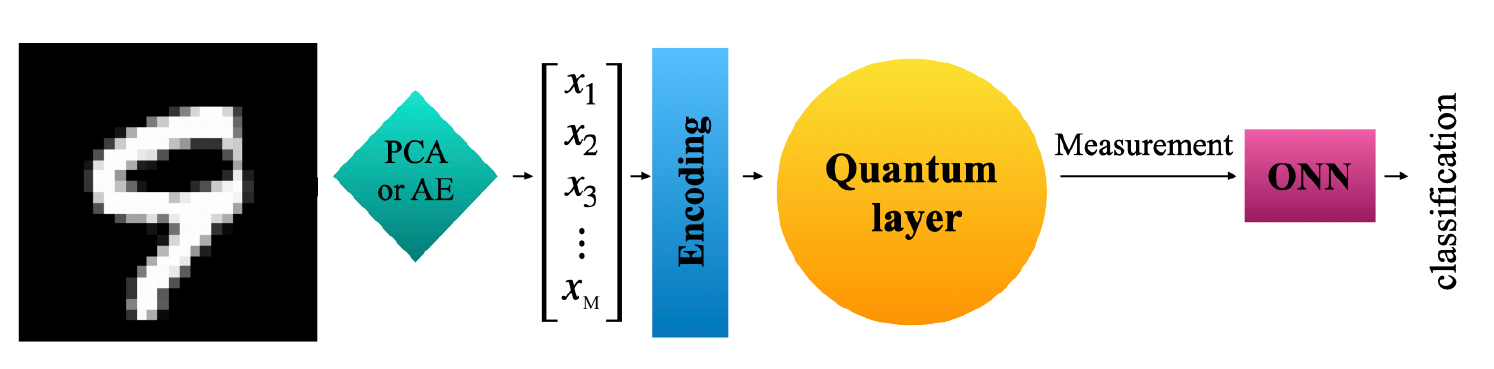}
    \caption{A schematic representation of the QELM. The workflow is as follows: feature reduction through PCA or AE; encoding of classical data into  the quantum initial state; time evolution via the quantum layer; measurement of the evolved quantum state; classification with a classical One-layer Neural Network.}
    \label{fig:QELM model}
\end{figure*}

We will explore a QELM model that combines a classical Extreme Learning Machine framework, in which the training process is restricted to the output layer only, with a quantum reservoir. As said before, we feed the algorithm with an image dataset (MNIST) to implement an image classification task. A schematic representation of the model is shown in Fig.\ref{fig:QELM model} and its workflow can be summarized as follows:
\begin{itemize}
  \item Feature reduction: the original data is compressed into a lower dimensional latent representation to match the limited number of available qubits of the reservoir. This can be achieved using either linear Principal Component Analysis (PCA) or a nonlinear Auto-Encoder (AE) \cite{KRAMER1992313,Hinton1993AutoencodersMD,Rumelhart1986LearningRB, doi:10.1126/science.1127647}.
  \item Encoding: the process involves transforming classical information into quantum states that can be processed by the quantum system. This entails setting the initial states of the qubits to be used for the subsequent quantum computation.
  \item Quantum layer and time evolution: the quantum system evolves in time according to the dynamics dictated by the Hamiltonian.
  \item Measurement: the qubits of the system are measured by performing projective measurements associated with specific operators, such as Pauli operators, or by executing a quantum state tomography procedure requiring multiple runs.
  \item Classical classifier: the output of the measurement process is fed into a simple One-layer Neural Network (ONN) that is trained to perform classification. 
\end{itemize}
\vspace{-12pt}
\section{Techniques for reducing data complexity}\label{sec:Principal Component Analysis vs Auto-Encoders}
Quantum machine learning holds promising potential. However, one of the primary challenges in applying this technology is the limited availability of qubits.
Additionally, most datasets of interest have a high number of features, necessitating the development of strategies and algorithms to reduce the dimensionality while preserving as much information as possible of the original data, see for instance \cite{GONG2024129993}.\\
In depth, we examine two distinct approaches for reducing the number of dimensions of the features: the Principal Component Analysis (PCA) and the Auto-Encoder (AE). \\
While the PCA is a statistical technique that transforms a set of potentially correlated variables into a subset of linearly uncorrelated ones, called principal components, selected according to their variance, the AE is a type of artificial neural network. The latter is trained to map input data to itself and it comprises two main components: the encoder, which compresses the input into a lower-dimensional latent-space representation, and the decoder, which reconstructs the input data from the latent-space representation with the aim to make the output as close to the original input as possible.\\
In particular, we used two different algorithms of auto-encoders that we name $AE_1$ and $AE_2$. Both utilize in the encoder convolutional and max-pooling layers for feature extraction, while convolutional and up-sampling layers are used in the decoder for the reconstruction of the input data. A fully-connected layer maps the aforementioned features into a latent-space representation which is then fed to the quantum algorithm. The fully-connected layer employs a sigmoid activation function (the outputs of which are real values between 0 and 1) while the convolutional layers rely on the rectified linear unit. The two auto-encoders mainly differ in their expressivity potential, with $AE_1$ characterized by a number of trainable parameters larger than the one in $AE_2$. More details on the architectures of $AE_1$ and $AE_2$ can be found in Appendix A.
\section{Encoding strategies of classical data into the QELM}\label{sec:Different encodings}
First of all, we remind the reader that the pure state of a qubit can be represented as a point on the Bloch sphere
\begin{equation} \label{eqn:blochsphere}
    \ket{\psi} = \cos (\frac{\theta}{2})\ket{0} + e^{ i \phi} \sin (\frac{\theta}{2})\ket{1}
\end{equation}
where $\theta \in [0,\pi]$ and $\phi \in [0,2\pi]$, to which we will refer when we will describe the various encodings. Encodings based on this representation can be easily used to transform classical data into quantum states, through a mapping of the classical features into the angles $(\theta, \phi)$ describing the position of a qubit on the Bloch sphere.
\subsubsection*{Dense angle encoding}
The representation defined above can be fully exploited to map two features per qubit using the polar angle and the relative phase between the two quantum states of each qubit. This is also dubbed “\textit{dense angle encoding}” and it is explicitly given by \cite{LaRose:2020mgo}
\begin{equation} \label{eqn:denseangleencoding}
    \ket{\vec{x}} = \bigotimes_{i=1}^{\ceil*{M / 2}} \left( \cos (\frac{x_{2i -1}}{2})\ket{0} + e^{ i x_{2i}} \sin(\frac{x_{2i -1}}{2})\ket{1}\right) 
\end{equation}
where $x_i$ represents the elements of the feature vector $\vec{x} = [x_1, ..., x_M]^T \in \mathbb{R}^M$ and $N = \ceil*{M / 2}$ is the number of qubits. We chose to normalize each feature in the interval $[0, \pi]$, thus effectively covering, for each qubit, half of the corresponding Bloch sphere \cite{Sakurai:2022ala}, with the angles $\theta$ and $\phi$ identified with pairs of features ($x_{2i -1}$, $x_{2i}$).
\subsubsection*{Angle encoding}
A simpler case is represented by a single qubit encoding which maps each feature into one qubit ($M = N$), without exploiting the relative phase, namely with $\phi = 0$. In this case, the initial state is prepared according to
\begin{equation} \label{eqn:qubit_encoding_grant}
   \ket{\vec{x}} = \bigotimes_{i=1}^{M} \left(\cos (\frac{x_i}{2})\ket{0} + \sin (\frac{x_i}{2})\ket{1}\right).
\end{equation}
This mapping is also called “\textit{angle encoding}” and has been used, for instance, in references \cite{LaRose:2020mgo, Schuld:2018gao, Grant:2018oml, Stoudenmire2016SupervisedLW, PhysRevA.101.052309}.
\subsubsection*{Uniform Bloch sphere encoding}
Another similar encoding, called “\textit{uniform Bloch sphere encoding}”, is given by the following representation \cite{Mujal_2021}
\begin{equation} \label{eqn:uniform_bloch}
    \ket{\vec{x}} = \bigotimes_{i=1}^{\ceil*{M / 2}} \left(\sqrt{x_{2i-1}}\ket{0} + e^{ i x_{2i}} \sqrt{1-x_{2i-1}}\ket{1}\right)
\end{equation}
where the features $x_{2i-1}$ are normalized between $[0, 1]$, while the $x_{2i}$, representing the phase $\phi$, must be normalized in the interval $[0, \pi]$, as explained above.
\subsubsection*{General encoding}
Among the so-called “\textit{general encodings}” \cite{LaRose:2020mgo}, we further consider
\begin{equation} \label{eqn:generalencoding}
    \ket{\vec{x}} = \bigotimes_{i=1}^{\ceil*{M / 2}} \frac{1}{\sqrt{x_{2i -1}^2+x_{2i}^2}} \left(x_{2i -1} \ket{0} + x_{2i} \ket{1}\right).
\end{equation}
Here the pairs of features ($x_{2i -1}$, $x_{2i}$) take values between $[0, 1]$.
\subsubsection*{Amplitude encoding}
Finally, we considered the “\textit{amplitude encoding}” \cite{LaRose:2020mgo}
\begin{equation} \label{eqn:amplitudeencoding}
    \ket{\vec{x}} = \frac{1}{||\vec{x}||_2} \sum_{i=1}^{2^N} x_{i}\ket{i} 
\end{equation}
where $x_i$ is the $i$th feature of $\vec{x}$ and $\ket{i} $ represents a vector of the computational basis. Although this encoding is difficult to achieve as it would require the ability to prepare a massively entangled state, it has the advantage that it can reduce the demand for qubits.
\section{Hamiltonian models describing the quantum layer}\label{sec:Different Hamiltonians}
In this section, with the aim of understanding the features of the dynamics that help to improve the performances, we explore the influence of various Hamiltonians generating qualitatively different dynamics of the qubit register. Specifically, we analyze six distinct Hamiltonians, which govern the time evolution of the initial state containing the encoded features,  on the classification accuracy. In particular, we consider the time-dependent Hamiltonian suggested in \cite{Sakurai:2022ala}, alongside five time-independent operators. Among the latter, two have the qubit lying on a fully connected graph, with all-to-all interactions, while three others model qubits arranged in a linear configuration, with only nearest neighbor interactions.\\
In the first case, the quantum layer performs the temporal evolution of a quantum state under the periodic time-dependent Hamiltonian $H_1$ \cite{Sakurai:2022ala}
\begin{equation}
\label{eqn:H1}
H_1(t)= \begin{cases} H_a= B_1 \sum_{i=0}^{N-1} \sigma_x^{(i)} \, \, \, \, \, \,  0\leq t<T_1 \\ H_b=\sum_{i,j=0}^{N-1} J_1^{i,j} \sigma_z^{(i)}\sigma_z^{(j)} \, \, \, \, \, \, T_1 \leq t<T \\ \end{cases}
\end{equation}
Here, $\sigma_x^{(i)}$, $\sigma_z^{(i)}$, and $\sigma_z^{(j)}$ are the Pauli operators acting on the site of the $i$,$j$-th qubit, while $J_1^{i,j}=\frac{J_0} {|i-j|^{\alpha}}$ represents the long-range interaction between the $i$-th and $j$-th qubits. As in \cite{Sakurai:2022ala}, the scale of the interaction is set by $J_0=0.06$ and is damped for far-away qubits, with $\alpha=1.51$ controlling the rate of damping. $B_1=3.05$ and the time evolution is performed using a $\Delta t = 50 T$ ($T=2T_1=1$, setting the units for both time and inverse energy).
\\
In the second case, we use the Hamiltonian $H_2$ below, featuring interacting qubits in a transverse field, \cite{Domingo:2022fre} 
\begin{equation}
     H_2= \sum_{i,j=0}^{N-1} J_2^{i,j} \sigma_z^{(i)}\sigma_z^{(j)} + \sum_{i=0}^{N-1} B_2^{(i)} \sigma_x^{(i)}.
\end{equation}
Here, the coefficients $J_2^{i,j}$ and $B_2^{(i)}$ are sampled from the Gaussian distributions $N(0.75, 0.1)$ and $N(1, 0.1)$, respectively. The time evolution is implemented with a time $\Delta t = 20$. \cite{Domingo:2022fre}\\
In the third case, we consider a Ising-like model, with only nearest neighbor interaction, including both longitudinal and transverse fields, as used in \cite{Xiong:2023qqh},
\begin{equation}
 H_3= J_3 \sum_{i=0} ^{N-2}\sigma_z^{(i)}\sigma_z^{(i+1)} + B_{3z}\sum_{i=0}^{N-1} \sigma_z^{(i)} + B_{3x}\sum_{i=0}^{N-1} \sigma_x^{(i)} 
\end{equation}
\\
where we set the parameters in the chaotic regime: $J_3=-1$, $B_{3x}=0.7$ and $B_{3z}=1.5$. For $H_3$ as well, the time evolution is performed using a $\Delta t = 20$\cite{Xiong:2023qqh}.\\
As a variation of the latter and as our fourth example, we employ the Heisenberg XXZ model, whose Hamiltonian we dub $H_4$:
\begin{align}
     H_4 = -\frac{1}{2} \sum_{i=1}^{N} \Big( 
     & J_{4x} \sigma^{(i)}_x \sigma^{(i+1)}_x 
     + J_{4y} \sigma^{(i)}_y  \sigma^{(i+1)}_y  \notag \\
     & + J_{4z} \sigma^{(i)}_z \sigma^{(i+1)}_z 
     + B_{4z} \sigma^{(i)}_z \Big)
\end{align}
where we set the parameters $J_{4x}=J_{4y}=2$ and $J_{4z}=B_{4z}=0.54$ and the time evolution is performed using a $\Delta t = 20$.\\
In Ref. \cite{Martínez2021}, the performance of QRC has been related to the localization and ergodicity properties of the Hamiltonian eigenspectrum. With the aim of testing the role of locality/ergodicity (and also of integrability) on the algorithm performance, we further explore two other kind of Hamiltonians; namely:  a fully integrable model and the Hamiltonian reported in Ref. \cite{Martínez2021}, which, depending on the values of its parameters, can display different localization features.\\
Thus, we consider the integrable and non-ergodic XX model, whose Hamiltonian describes (in the fermion language) just hopping between nearest neighbours, $H_5$:
\begin{equation}
 H_5= \frac{1}{2} \sum_{i=1}^{N} (\sigma_x^{(i)}\sigma_x^{(i+1)} +\sigma_y^{(i)}\sigma_y^{(i+1)}).
\end{equation}
As our final example, we take the Hamiltonian $H_6$ reported in Ref. \cite{Martínez2021}:
\begin{equation}
 H_6= \sum_{i>j=1}^{N} J_6^{i,j} \sigma^{(i)}_{x}\sigma^{(j)}_{x} + \frac{1}{2} \sum_{i=1}^{N} (B_6+D_i) \sigma^{(i)}_{z}
\end{equation}
where $J_6^{i,j}$ are randomly selected from a uniform distribution in the interval $[-0.5, 0.5]$, $D_i$ is randomly drawn from the uniform distribution $[-W, W]$. In detail, we analyzed the reservoir dynamics generated by $H_6$ using three different set of parameters, corresponding to non-ergodic regions described in \cite{Martínez2021}: a first one, where localization occurs (region III in Ref. \cite{Martínez2021}) with $B_6=W=2*10^{-2}$, a second one, laying in between non-ergodic and ergodic regions, with $B_6=0.03$ and $W=1$ and a third set of parameters lying in the many-body-localized region (region I of Ref. \cite{Martínez2021}), with $B_6=0.03$ and $W=60$.\\
Moreover, to help the comparison we have also taken into account the Hamiltonian $H_2$ with $J_2=0$, as this choice represents a system where there are no interactions between the spins.\\
The evolution generated by all of these Hamiltonians, implemented using the QuTiP package \cite{JOHANSSON20131234, JOHANSSON20121760}, is taken for a time $\Delta t = 20$.
\section{Measurement and machine learning classifier}\label{sec:ML classifier}
After the quantum system evolves in time, following the dynamics governed by the chosen Hamiltonian, we measure the output state in the computational basis. This allow us to extract the classical information consisting in the population density of the wave function of the final state. In a real experiment, we can statistically gather this data by repeating the procedure multiple times. In this way we obtain $2^N$ real values for each image which will be then fed to the machine learning classifier. The latter consists of a fully-connected single-layer neural network with a softmax activation function. The structure of the layer is completely fixed by the number of qubits used in the quantum layer and by the number of different classes in the input dataset. The model is trained by minimising the categorical cross entropy using the Adam optimiser. \\
For the numerical implementation, we used the Keras framework \cite{chollet2015keras}.
\section{Results}\label{sec:Results}
We defined as baseline configuration the case of using autoencoder $AE_1$, dense angle encoding, and Hamiltonian $H_1$. Afterward, we varied each component of the workflow individually while keeping the baseline choice for the other components and studied the performance in the form of the achieved accuracy as a function of a number of qubits or features. Due to the computational resources needed to simulate quantum computers with a large number of qubits, we had to limit the study to systems with up to 12 qubits. The findings are summarized in the following subsections.
\subsection{PCA vs Autoencoders}
\begin{figure*}[htbp]
    \centering
    \subfigure[]{
        \includegraphics[width=0.45\textwidth]{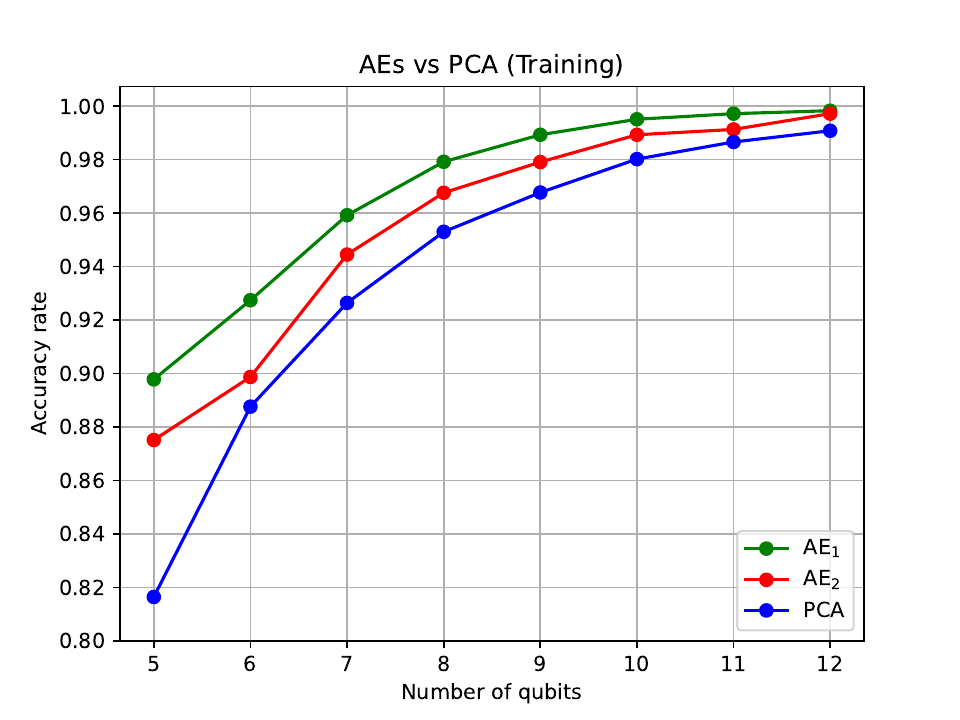}
    }
\hspace{0.001\textwidth}
    \subfigure[]{
        \includegraphics[width=0.45\textwidth]{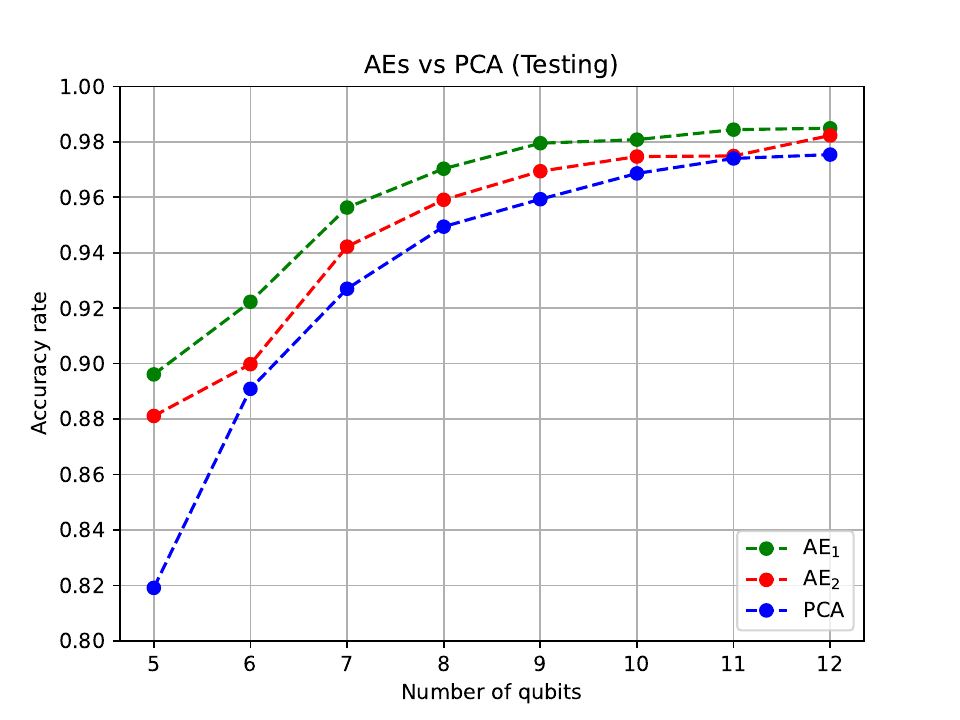}
    }
    \caption{Training (left panel) and testing (right panel) accuracy as a function of the number of qubits for different feature reduction schemes: PCA (blue) and two autoencoders ($AE_1$ and $AE_2$, green and red
curves respectively). The  time evolution and the encoding have been performed with the Hamiltonian $H_1$ and the dense angle, respectively.}
    \label{fig:PCA vs AEs}
\end{figure*}

In the two plots in Fig. \ref{fig:PCA vs AEs} we show the accuracy evaluated on the training and test sets as a function of the number of qubits of the quantum layer for three different setups. \\
In particular, we have explored the architecture performance with respect to the feature-reduction strategies, namely, a PCA and two autoencoders. We fixed, instead, the encoding algorithm to the “dense angle encoding”, as detailed in Eq. \ref{eqn:denseangleencoding}, and the Hamiltonian to the time-dependent $H_1$ in Eq. \ref{eqn:H1}. The dimension of the latent space scales as $2N$ as shown in Tab. \ref{tab:Encodings}.\\
As naively expected, the simpler linear PCA shows a lower accuracy compared to the neural-network-based algorithms, $AE_1$ and $AE_2$, and $AE_1$ performs better than $AE_2$ due to its higher capacity. We notice that, after a fast initial growth, the accuracy starts to saturate around $N = 9$ qubits to $\sim 98 \%$ for the test set and slightly higher for the training set. As the number of qubits increases, the differences among the classification performances of the three algorithms get reduced. This behaviour can be ascribed to the corresponding increase in the dimension of the latent representation, combined with the relative simplicity of the MNIST dataset. As such, we expect the gap between the autoencoders and the PCA to persist even for large $N$ for more complex datasets. As an example, we considered the Fashion-MNIST dataset and confirmed the presence of the gap. The corresponding results are provided in Appendix B.
\subsection{Encodings}
\begin{figure*}[htbp]
    \centering
    \subfigure[]{
        \includegraphics[width=0.45\textwidth]{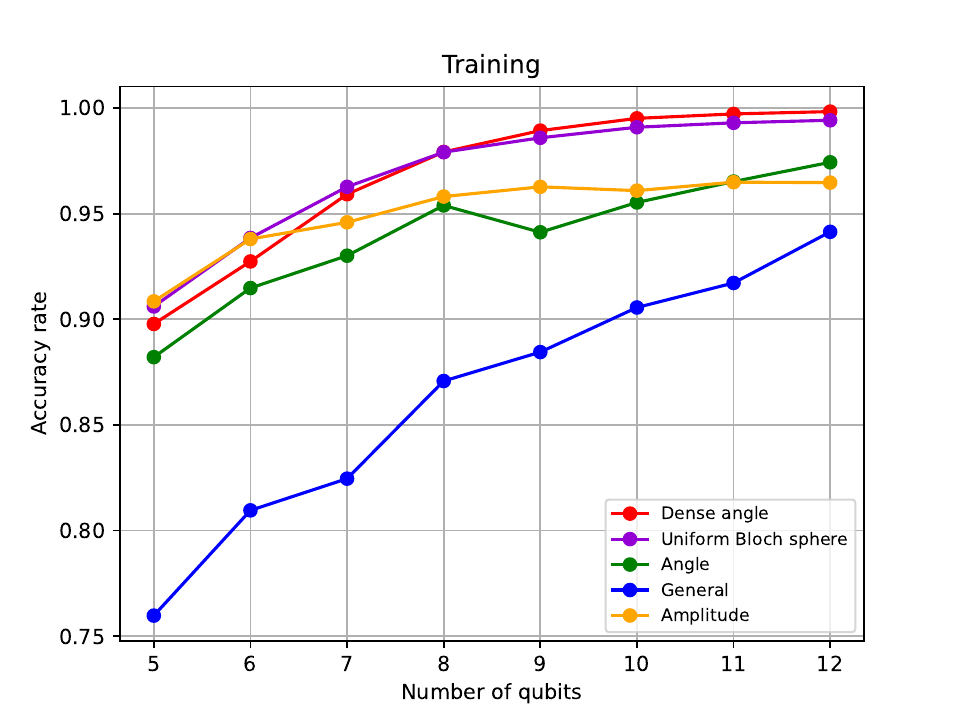}
    }
     \hspace{0.001\textwidth}
    \subfigure[]{
        \includegraphics[width=0.45\textwidth]{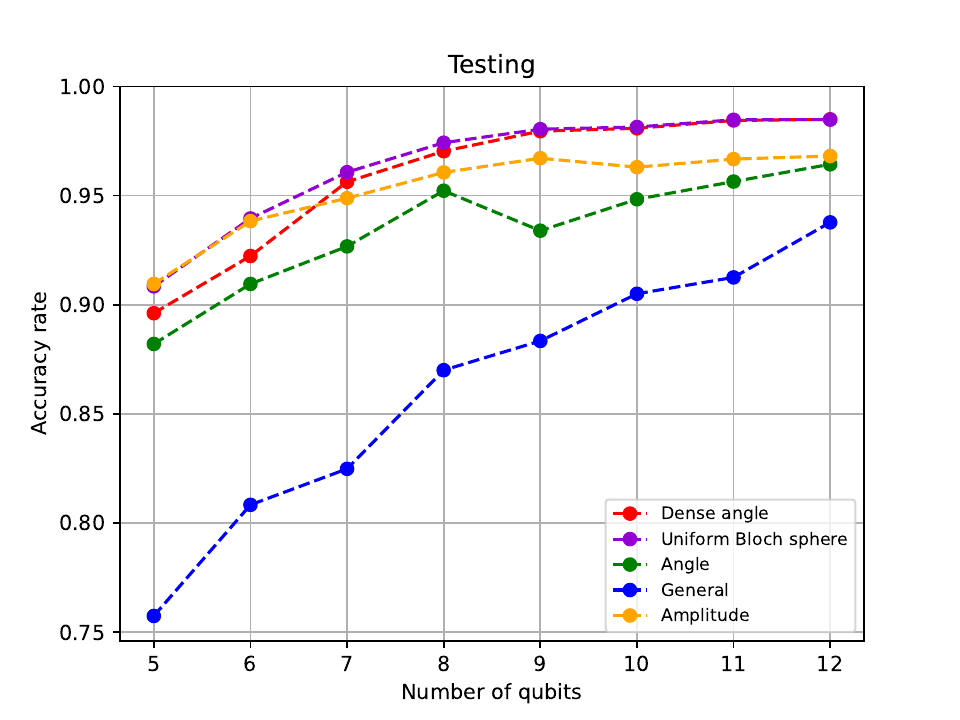}
    }
    \caption{Training (left panel) and testing (right panel) accuracy as a function of the number of qubits for different encoding strategies: dense angle (red), uniform Bloch sphere (purple), angle (green), general (blue) and amplitude (orange) encodings. The feature reduction has been performed with the autoencoder $AE_1$ and the time evolution with the Hamiltonian $H_1$.}
    \label{fig:Different encodings to be completed}
\end{figure*}

\begin{figure*}[htbp]
    \centering
    \subfigure[]{
        \includegraphics[width=0.45\textwidth]{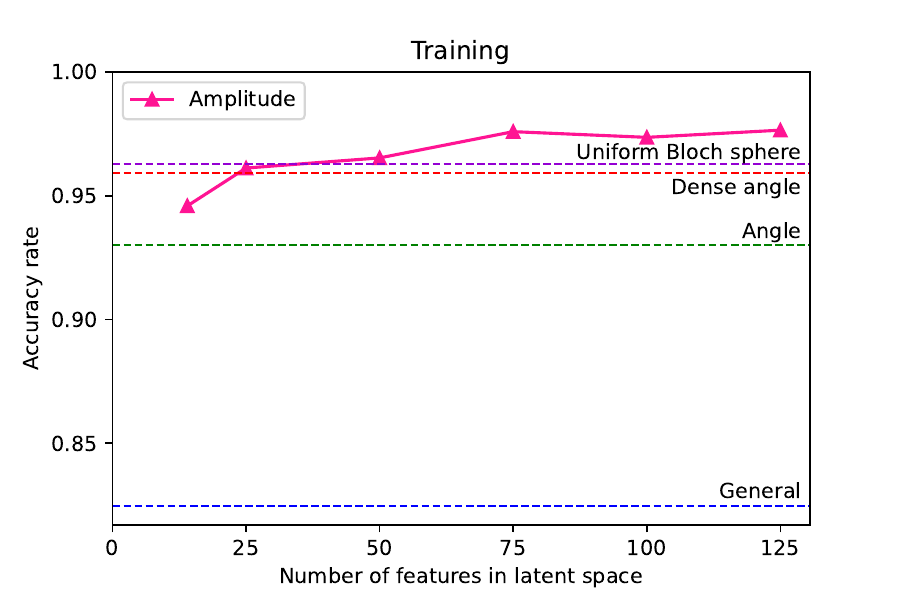}
    }
     \hspace{0.001\textwidth}
    \subfigure[]{
        \includegraphics[width=0.45\textwidth]{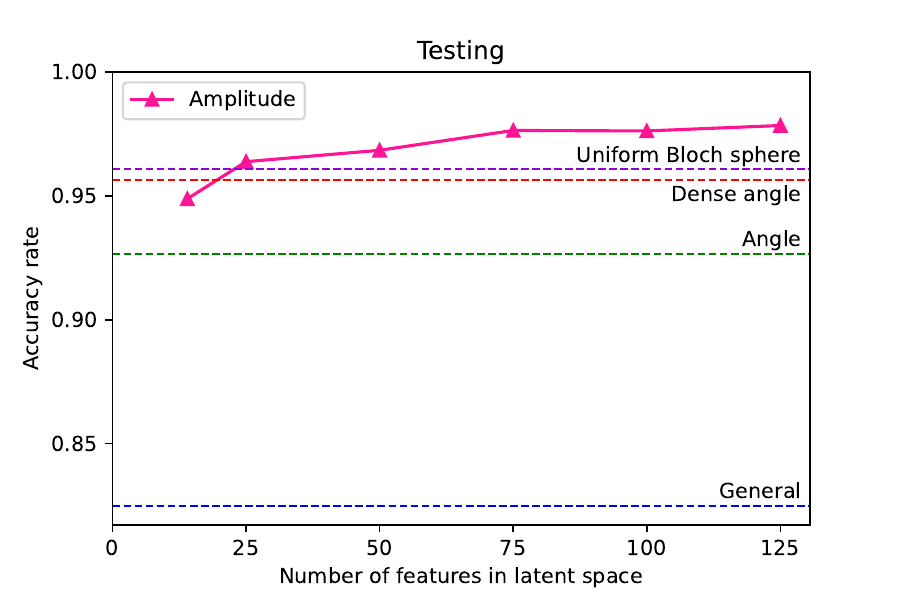}
    }
    \caption{Training (left panel) and testing (right panel) accuracy as a function of the number of features in the latent space obtained with the amplitude encoding with $N=7$. The feature reduction has been performed with the autoencoder $AE_1$ and the time evolution with the Hamiltonian $H_1$. Dashed horizontal lines represent the accuracy achieved with the other encoding schemes with the same number of qubits, $N=7$.}
    \label{fig:Amplitude encoding}
\end{figure*}

In Fig. \ref{fig:Different encodings to be completed} we explore the discriminating power of the quantum classifier for the different encodings explained in Sec. \ref{sec:Different encodings}, namely, the dense angle encoding (red line), the uniform Bloch sphere encoding (purple line) as well as the angle (green), general (blue) and amplitude (orange) encodings. For this analysis, we used, as baseline, the $AE_1$ autoencoder and the $H_1$ Hamiltonian, and we trained the algorithm using an increasing number of qubits, from 5 up to 12. \\
We recall that for all the explored encodings but the amplitude encoding, the mapping of the features is performed separately on each qubit, one feature per qubit in the “angle encoding” case and two features per qubit in all the other cases.\\
For the amplitude encoding, instead, the features are spanned over the computational basis, allowing us to encode up to $2^N-1$ features. However for the sake of comparison, the dimension of latent space in the autoencoder is set to $2 N$, see Tab. 1 for details.\\ 
It is clear from both the training and testing accuracy, on left and right panels in the plots, respectively, that the dense angle and the uniform Bloch sphere encodings provide comparably the best performance over all the setups, especially for larger quantum reservoirs. The main implementation difference between the two resides in how the features are distributed along the polar direction in the Bloch sphere.\\
Among the “angle-like” encoding strategies, there is a clear accuracy gap between the two encodings that spread the features over the two angle directions of the Bloch sphere (two features per qubit) and the one in which only one angle is employed (one feature per qubit).\\
The worst performance is achieved by the general encoding setup, which is particularly poor for small quantum layers. This is expected since, by construction, pairs of different but proportional features are mapped to equivalent quantum states.\\
In the amplitude encoding case, the initial state is prepared by mapping the features onto the computational basis, thus taking advantage of a bigger portion of the Hilbert space. For this reason, one can afford much larger latent spaces (or, equivalently, reduce the demand of qubits) that can be exploited to better identify the different classes in the original dataset. As such, one expects the amplitude encoding to outperform the other methods for a sufficiently large number of features and fixed and small number of qubits.
We show in Fig.\ref{fig:Amplitude encoding} the accuracy of the quantum classifier with the amplitude encoding as a function of the dimension of the latent space. For definiteness, we fix the number of qubits to $N=7$ and we show, for comparison, the performances achieved with the other encodings.
\begin{table*}[htbp]
\centering
\begin{tabular}{c||cccccccc}
\toprule
\toprule
\diagbox{Encoding}{N° of qubits} & 5 & 6 & 7 & 8 & 9 & 10 & 11 & 12\\
\midrule
\midrule
Angle & 5 & 6 & 7 & 8 & 9 & 10 & 11 & 12\\
Dense angle & 10 & 12 & 14 & 16 & 18 & 20 & 22 & 24\\
Uniform Bloch sphere & 10 & 12 & 14 & 16 & 18 & 20 & 22 & 24\\
General & 10 & 12 & 14 & 16 & 18 & 20 & 22 & 24\\
Amplitude & 10 & 12 & 14 & 16 & 18 & 20 & 22 & 24\\
\bottomrule
\bottomrule
\end{tabular}
\caption{Dimension of the latent space for different number of qubits and encodings.
}
\label{tab:Encodings}
\end{table*}
\vspace{-10pt}
\subsection{Hamiltonians}
\begin{figure*}[htbp]
    \centering
    \subfigure[]{
        \includegraphics[width=0.45\textwidth]{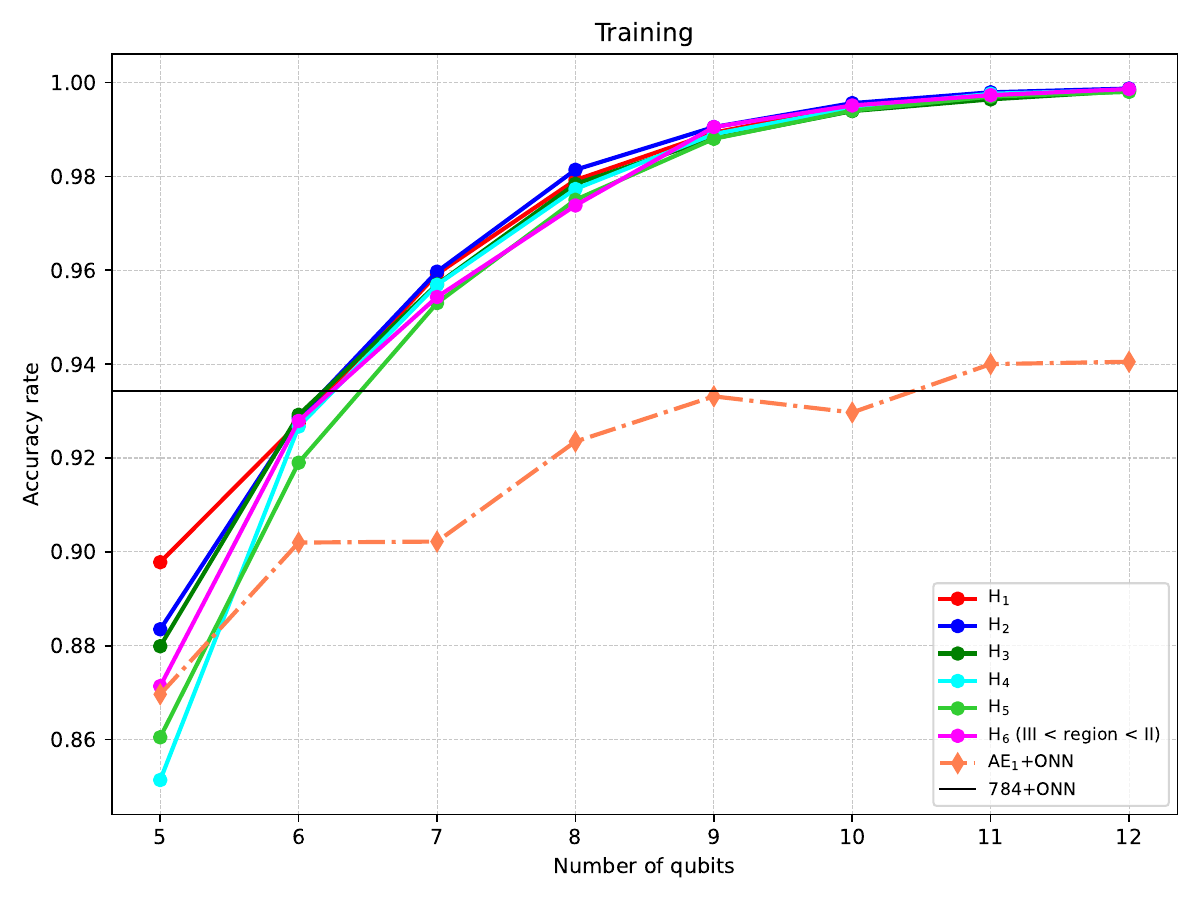}
    }
     \hspace{0.001\textwidth}
    \subfigure[]{
        \includegraphics[width=0.45\textwidth]{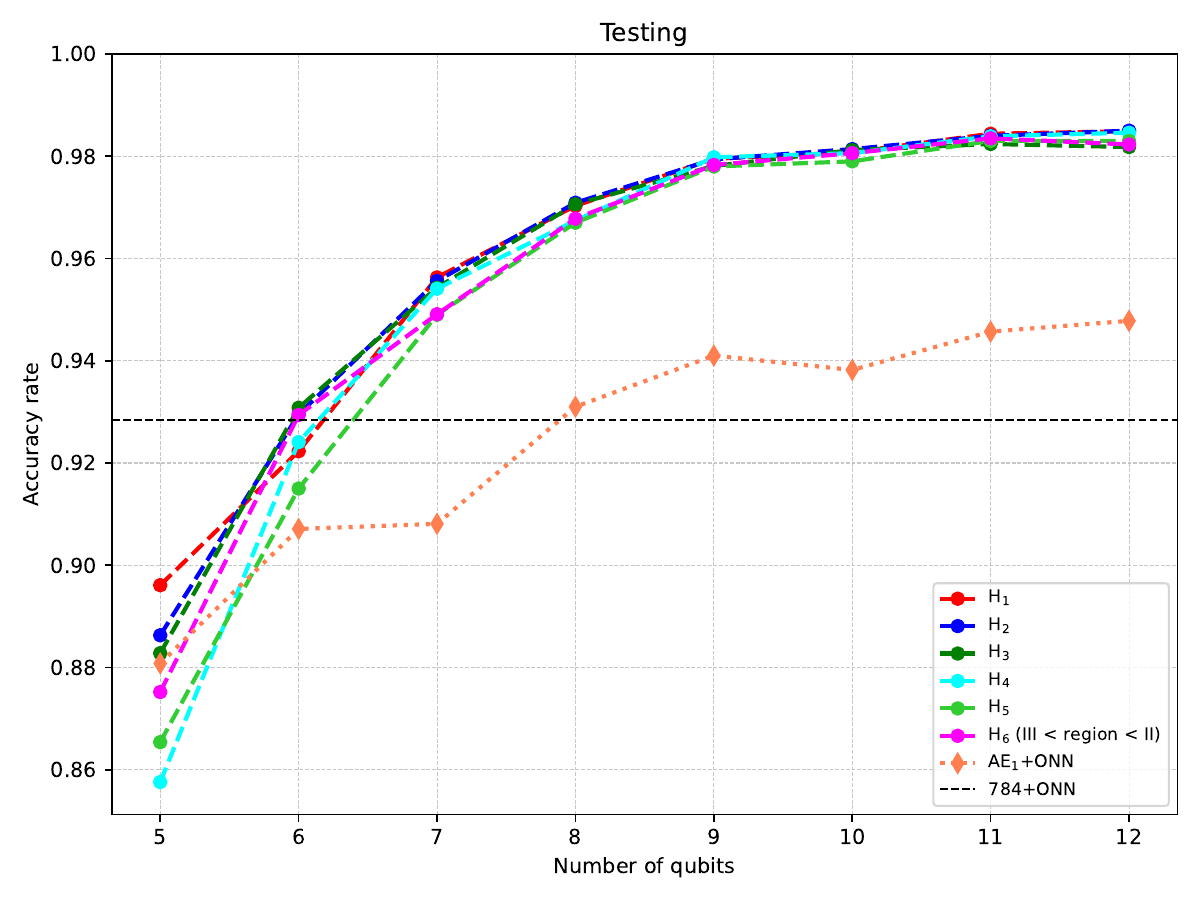}
    }
    \caption{Training (left panel) and testing (right panel) accuracy as a function of the number of qubits for six different Hamiltonians. The feature reduction and the encoding have been performed with the autoencoder $AE_1$ and the dense angle encoding, respectively. The horizontal line represents the accuracy obtained with a ONN fed with all the pixel values of the images. Instead, the coral line represents a ONN fed directly with the $2N$ features extracted by the $AE_1$. }
    \label{fig:Different Hamiltonians}
\end{figure*}

\begin{figure*}[htbp]
    \centering
    \subfigure[]{
        \includegraphics[width=0.45\textwidth]{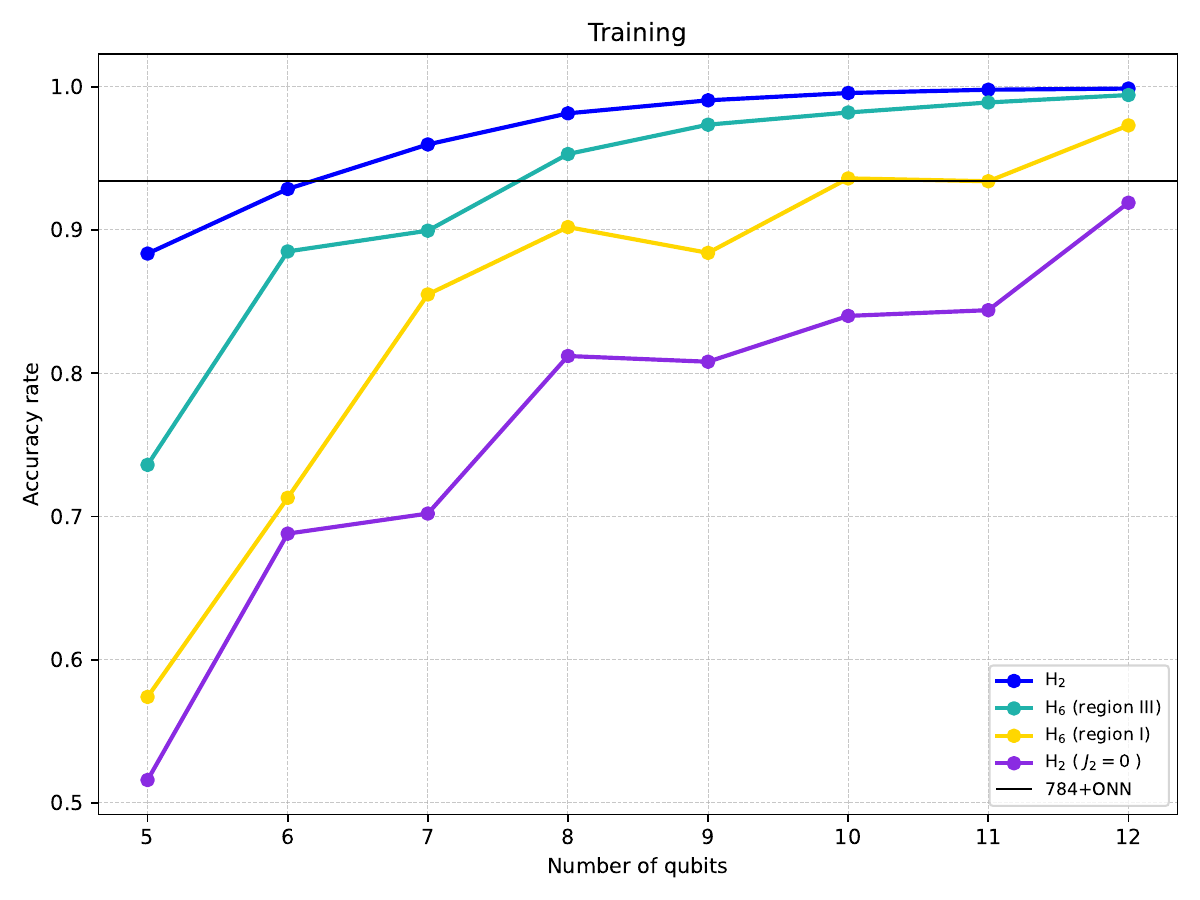}
    }
     \hspace{0.001\textwidth}
    \subfigure[]{
        \includegraphics[width=0.45\textwidth]{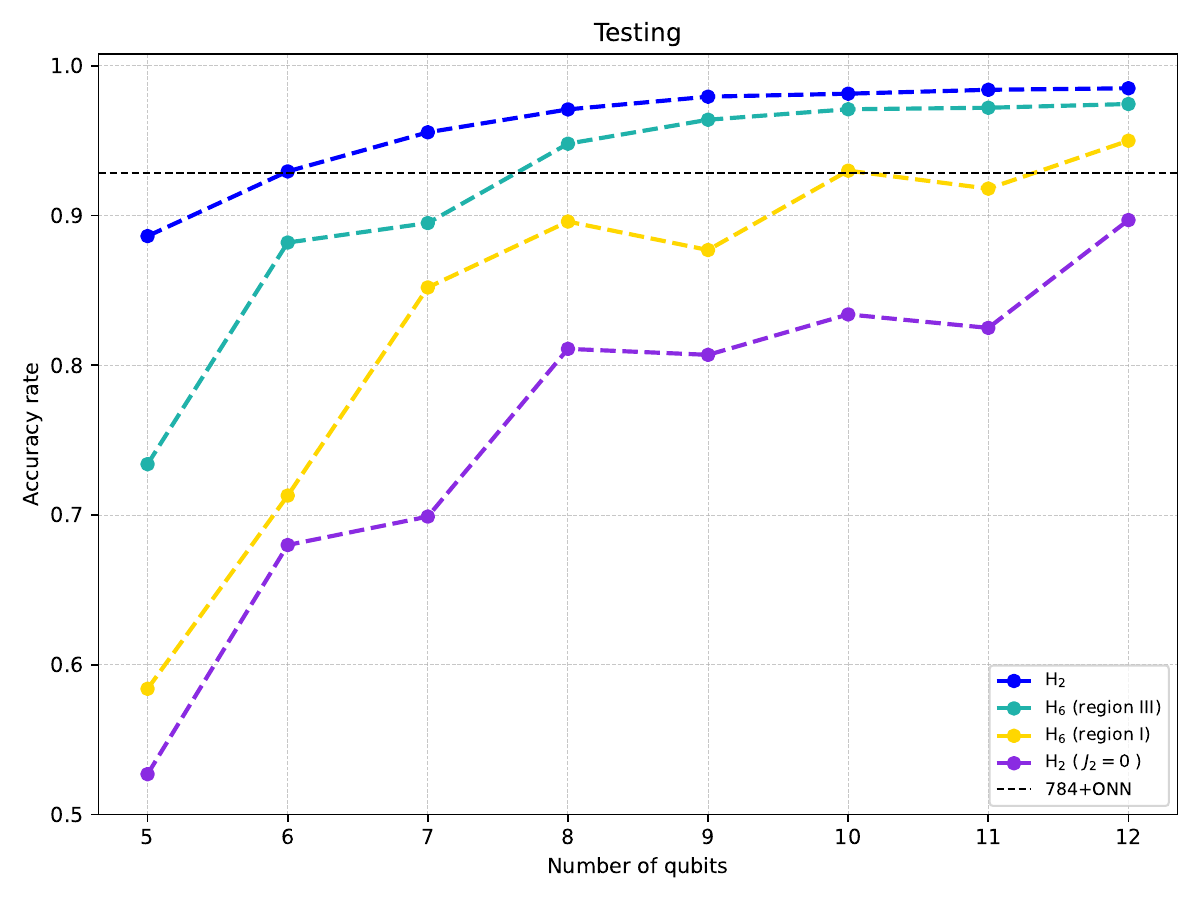}
    }
    \caption{Training (left panel) and testing (right panel) accuracy as a function of the number of qubits for other different Hamiltonians. The feature reduction and the encoding have been performed with the autoencoder $AE_1$ and the dense angle encoding, respectively. The horizontal line represents the accuracy obtained with a ONN fed with all the pixel values of the images.}
    \label{fig:Different Hamiltonians_bad}
\end{figure*}

Finally, we fix the feature reduction strategy and the encoding method and investigate the performance of the quantum classifier by comparing the different Hamiltonians introduced in Sec.\ref{sec:Different Hamiltonians}. In detail, we can see in Fig.\ref{fig:Different Hamiltonians} similar and good accuracy performance as function of the number of qubits for the time-dependent Hamiltonian  $H_1$, the transverse-field Ising Hamiltonians $H_2$ and $H_3$, the Heisenberg XXZ model $H_4$, the integrable and non-ergodic XX $H_5$ and the Hamiltonian $H_6$ with parameters in the non-ergodic and transition region. The accuracy rapidly grows with increasing the number of qubits and reaches a plateau around $N=10$.\\
This lack of sensitivity can be attributed to the complex nature of the dynamics generated by these Hamiltonians, it spreads the information from the local states (where the encoding occurs) to the computational basis, combining the different pieces of information encoded in different qubits. When the probability of the outcomes is measured, the recombined information is extracted, allowing the device to learn from the process.\\
This, however, does not occur, for instance, when Hamiltonians that induce localization are used. We show in Fig. \ref{fig:Different Hamiltonians_bad} the performance accuracy as function of the number of qubits for the Hamiltonians $H_6$ in localizations areas (region I and III of Ref. \cite{Martínez2021}) and for transverse-field Ising Hamiltonian $H_2$ when $J_2 = 0$, that is where there are no interactions among the spins. In the Fig. \ref{fig:Different Hamiltonians_bad}, we have included $H_2$ again, as a reference for the better performance of the Hamiltonians, and here we can notice how $H_6$ and $H_2 (J_2=0)$ have worse performance.\\
For comparison, in Fig. \ref{fig:Different Hamiltonians} and Fig. \ref{fig:Different Hamiltonians_bad}, we also 
show the classification accuracy obtained after stripping off the autoencoder and the quantum layer; namely, by a purely classical ONN which is fed with the full image information (the 784 pixel values). The classical and quantum classifiers achieve similar results for $N=6$, but the latter clearly outperforms if more qubits are employed.\\
Additionally, in Fig. \ref{fig:Different Hamiltonians} we present a further classical comparison with an ONN directly fed with the features reduced by the autoencoder $AE_1$ (with a latent space dimension given by $2N$). Also in this case, after $N=6$, the quantum classifier outperforms the classical one. \\
The accuracy obtained with the ONN using the raw 784 data is in general greater than, or comparable with, the one achieved with the compressed features (coral line), the latter slightly exceeding the former when $N = 11$ for the train set and $N = 8$ for the test set. This suggests that, for a straightforward comparison with the quantum model, the ONN with the raw data represents a simple and viable choice, at least for the range of $N$ analyzed in this work.\\
Furthermore, it is important to emphasize that the task of the autoencoder is not the classification itself, but rather the reduction of the amount of available information. While in some cases this may help to remove the noise, in general it leads to a loss of data. The use of the AE in the quantum model is not a strategic advantage over the classical model in the classification task, but rather a necessity due to the current limitations of quantum hardware, which can only handle a limited number of qubits.\\
To further investigate the role of the autoencoder and, more generally, of a compressed representation, one could devise in principle a quantum algorithm to which the raw and uncompressed data is directly provided as input. While this cannot be achieved using the QELM architectures developed in our work, due to the aforementioned computational limitations, the importance of the compressed representations, as compared to the uncompressed one, can still be highlighted. Indeed, we are able to encode the full data using the amplitude encoding, provided that $N \ge 10$ (the other encodings would require at least $N = 392$ qubits to store the entire image). In this case, the accuracy on the test sets, using $H_1$ as an example, is found to be $\simeq 0.968$ for $N=10$, $0.975$ for $N=11$ and $0.976$ for $N=12$. These accuracies are smaller than those obtained, using the same number of qubits ($N=10,11,12$), with the dense angle encoding, for which the compressed representation of the dataset is strictly necessary (see for instance Fig. \ref{fig:Different Hamiltonians}(b)).
Besides the actual performance of the algorithms, we want to emphasize once more, that the use of the compressed representation is way more efficient as it requires less resources while preserving as much information as possible of the original data. \\
Finally, Fig. \ref{fig:Different Hamiltonians}(a) and Fig. \ref{fig:Different Hamiltonians}(b) clearly show that the larger accuracies obtained with the quantum algorithm are not due to the dimensionality reduction, but to the presence of the quantum layer.

\section{Conclusions}\label{sec:conclusion}
In this work, we have explored several implementational aspects of quantum machine learning techniques for image classification problems. 
We went through the full quantum machine learning pipeline, from the preparation of the classical data to the final classification task.\\
To demonstrate our methodology in a concrete manner and identify the most promising approaches in a controlled setting, we employed the MNIST handwritten digits dataset for a multilabel classification problem. Moreover, our experiments with the Fashion-MNIST dataset confirm that this model can be readily extended to more complex and realistic scenarios.\\
The first step in the workflow consists of the reduction of the features of the original data into a compressed representation. 
This is an essential step to comply with the limited number of qubits available/simulable with the current technology. \\
We compared the final performance of the quantum machine learning algorithms applied to two different schemes for data reduction: one using a (linear) PCA and another exploiting 
non-linear autoencoders. As expected, the great expressive capabilities of a neural network-based architecture outperform the linear techniques, even for a simple dataset such as the MNIST.
We expect that the advantage of the usage of an autoencoder becomes even more critical for more complex datasets.\\
Classical data must be converted into quantum states that can be later processed by quantum circuits. This operation is usually dubbed “encoding”. \\
We inspected five different encoding strategies typically considered in the literature. Four of them operate on each qubit separately, while another one spans the whole Hilbert space. We found that, among the former, the dense angle and the uniform Bloch sphere encodings provide the best performance and, in contrast to the amplitude encoding, enjoy the easiest practical implementation on real quantum circuits. This result is particularly encouraging, as it naturally maximizes the trade-off between performance and feasibility.\\
On the other hand, although amplitude encoding may be difficult to realize on real hardware, it maps more features compared to the single-qubit methods and, therefore, can provide better accuracies if the number of qubits is small.\\
Finally, we studied the time evolution governed by six different Hamiltonians. The first one is characterized by an explicit time dependence, while the other five are time-independent. Of the latter, two are characterized by a system of qubits with all-to-all interactions, while others involve interactions effective only among closest neighbors. Moreover, among them, there are Hamiltonians we have tested to explore the role of ergodicity, integrability, and spin-spin interaction in the performance of the QELM.\\ 
We found that integrability and ergodicity do not play a distinguishing role in themselves; rather, it seems that a key factor is the spread of information from the initial local states to the final measurement states in the computational bases, passing through the eigenstates of the Hamiltonian. This does not work well, for example, in the case of localization.\\
Remarkably, the classification performance is effectively the same for the different Hamiltonians, thus suggesting that we could safely opt for the one that is easier to implement  without affecting the final gain.\\
For the final classification task, we used a single-layer fully-connected neural network, in line with the extreme learning machine prescription.\\
The present analysis could be extended in several directions. For instance, one could define several performance metrics (e.g., classification accuracy, time complexity, energy consumption, etc., and/or combinations of them) to compare the QELM algorithms against different classical models. This could be useful to quantify the quantum advantage. Moreover, one could explore the actual implementability of the models with real hardware.
\section*{Acknowledgments}
The results have received funds from the program Plan de Doctorados Industriales of the Research and Universities Department of the Catalan Government (2022 DI 011). The research leading to these results has received funding from the Spanish Ministry of Science and Innovation PID2022-136297NB-I00 /AEI/10.13039/501100011033/ FEDER, UE. IFAE is partially funded by the CERCA program of the Generalitat de Catalunya. JS acknowledges the contribution from PRIN (Progetti di Rilevante Interesse Nazionale) TURBIMECS - Turbulence in Mediterranean cyclonic events, grant n. 2022S3RSCT CUP H53D23001630006, CUP Master B53D23007500006. FP and NL aknowledge support by the PNRR MUR Project N. PE0000023-NQSTI.
\begin{appendices}
\section{Appendix}
\begin{figure*}[htbp]
    \centering
    \subfigure[AE$_1$]{
        \includegraphics[width=0.45\textwidth]{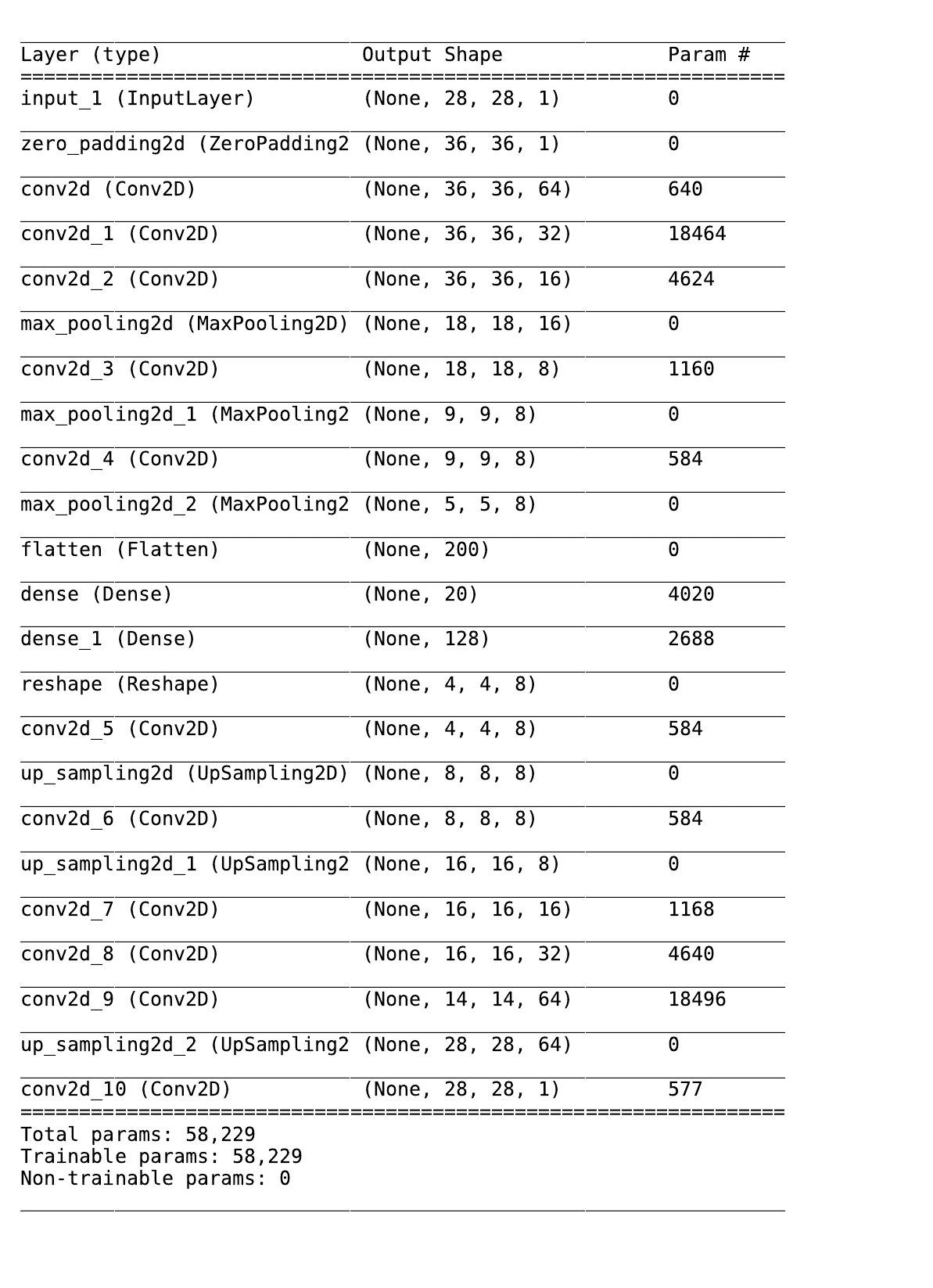}
    }   \hspace{0.001\textwidth}
    \subfigure[AE$_2$]{
        \includegraphics[width=0.45\textwidth]{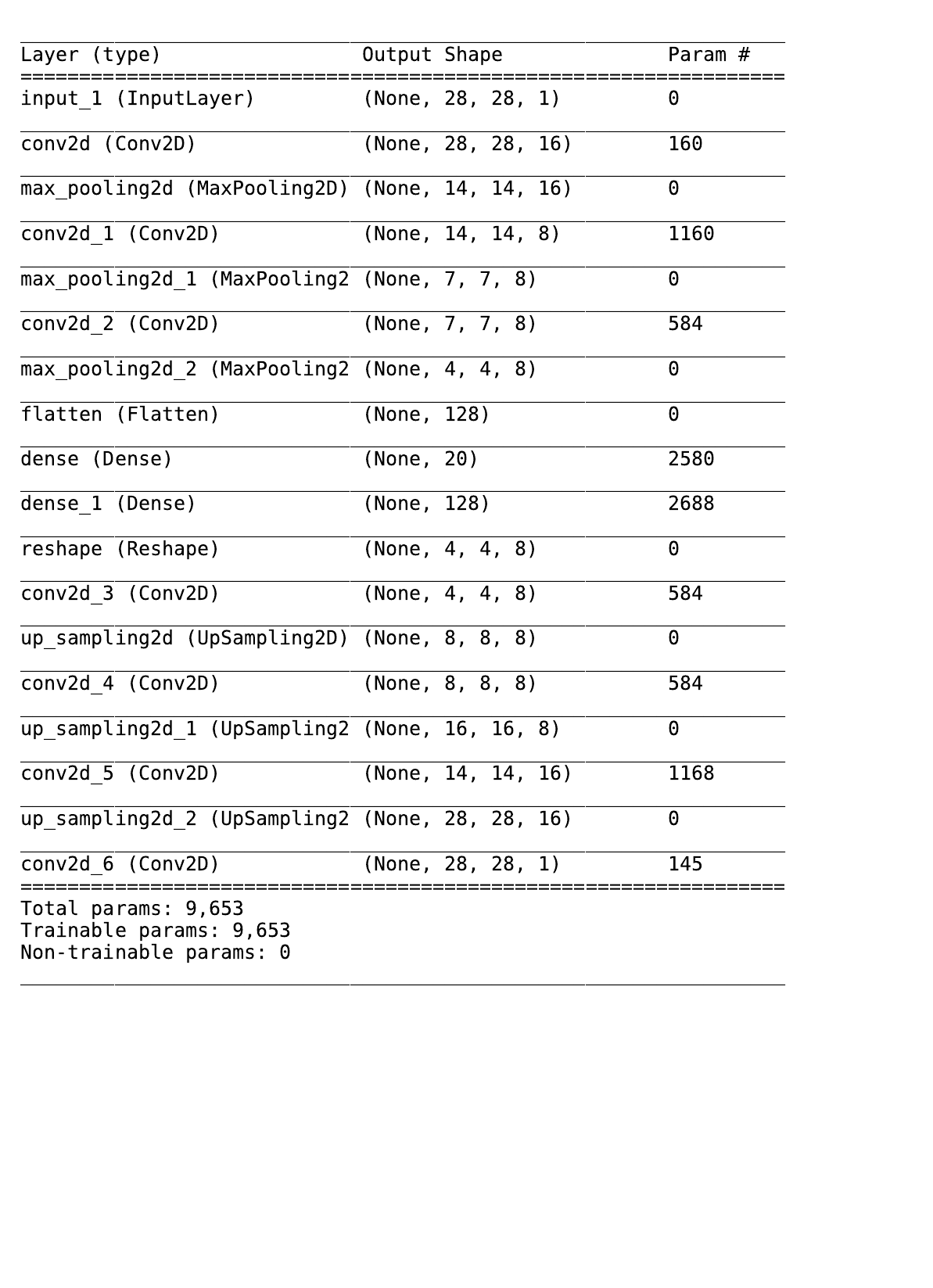}
    }
    \caption{Details of the architecture of the two autoencoders explored in the present analysis. }
    \label{fig:AEs}
\end{figure*}

In Fig. \ref{fig:AEs} we show the details of the architecture of the two autoencoders explored in this work. The main difference between them resides in the depth and the number of trainable parameters. The autoencoder consists of two parts, an encoder which compresses the input data into a lower-dimensional latent space representation, and a decoder, the purpose of which is to reconstruct the original information. The structure of the encoder is characterised by convolutional layers followed by max pooling layers which downsize the image during the feed-forward pass. The encoder ends with a fully-connected layer that ultimately maps the features into the latent space. The decoder, instead, is given by a sequence of convolutional layers and upsampling layers that gradually recover the original image size. 
For the activation functions of the convolutional layers we chose the rectified linear unit, while for the fully-connected layer in the last part of the encoder we employed a sigmoid function.\\
The autoencoders have been trained with the Adam optimiser for 50 epochs and with a binary-cross entropy for the loss function.
\section{Appendix}
\begin{figure*}[htbp]
    \centering
\includegraphics[width=0.55\textwidth]{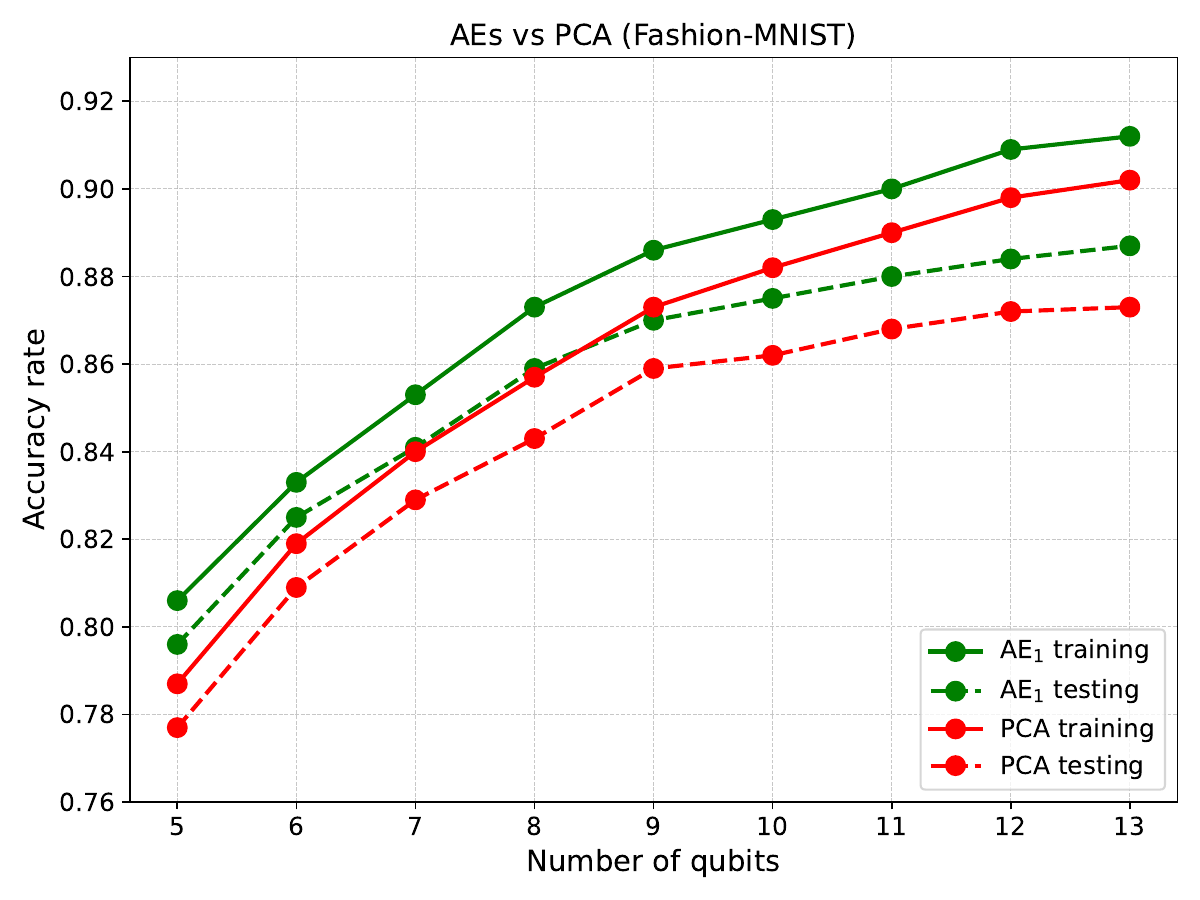}
    \caption{Training (solid line) and testing (dashed line) accuracy as a function of the number of qubits for different feature reduction schemes for the Fashion-MNIST dataset: PCA (red) and autoencoder $AE_1$ (green). The  time evolution and the encoding have been performed with the Hamiltonian $H_1$ and the dense angle encoding, respectively.}
    \label{fig:Fashion-MNIST}
\end{figure*}

To assess the robustness of our QELM algorithm, we also tested it on a more complex dataset, Fashion-MNIST. This dataset contains 70000 28x28 grayscale images of 10 different fashion categories, including clothing items such as t-shirts, shoes, and dresses.\\
We reduced the images using either the autoencoder $AE_1$ or PCA and fed them into our algorithm for image classification. The conclusions that can be drawn from the results in Fig. \ref{fig:Fashion-MNIST} are consistent with those found for the MNIST dataset and the gap between the accuracies of $AE_1$ (green line) and PCA (red line) persists in the range of qubits studied in this work.
\end{appendices}
\bibliographystyle{JHEP}
{\scriptsize
}
\providecommand{\href}[2]{#2}\begingroup\raggedright

\endgroup

\end{document}